\documentclass  [12pt] {amsart}
\usepackage{amssymb}
\usepackage{amsmath}
\usepackage{amsfonts}
\usepackage{graphicx}
\usepackage{epstopdf}
\usepackage[backref]{hyperref}
\usepackage{hyperref}

\usepackage{float}
\usepackage{placeins}
\makeatletter
\renewcommand\@biblabel[1]{#1.}
\makeatother

\newtheorem{proposition}{Proposition}

\theoremstyle{definition}
\newtheorem{remark}{Remark}
\newtheorem{example}{Example}[section]
\newtheorem{definition}{Definition}

\newcommand{\beq}{\begin{equation}}
\newcommand{\eeq}{\end{equation}}

\def\g{\gamma}
\def\G  {{\mathcal G}}
\def\b{\beta}
\def\p{\partial}
\def\a{\alpha}
\def\k{\kappa}
\def\g{\gamma}
\def\z{\zeta}

\def\gg {{\mathfrak g}}
\def\G {\Gamma}
\def\Re {{\rm Re\,}}
\def\l {\lambda}
\def\x {{\bf x}}
\def\y {{\bf y}}

\begin{document}

\title[Universal volume]{Universal volume of groups\\ and \\ anomaly of Vogel's symmetry }

            \author{H.M.Khudaverdian}
       \address{Manchester University and Max Planck Institute for Mathematics}
           \email{khudian@manchester.ac.uk}

       \author{R.L.Mkrtchyan}
       \address{Yerevan Physics Institute and Max Planck Institute for Mathematics}
           \email{mrl55@list.ru}

\maketitle

{\small  {\bf Abstract.} We show that integral representation of universal 
volume function of compact simple Lie groups gives rise to six analytic 
functions on $CP^2$, which transform as two triplets under group of 
permutations of Vogel's projective parameters.  This substitutes 
expected invariance under permutations of universal parameters by more 
complicated covariance. 

 We provide an analytical continuation of these functions and particularly 
calculate their change  under  permutations of parameters.  This last 
relation is universal generalization, for an arbitrary simple Lie group 
and an arbitrary point in Vogel's plane, of the Kinkelin's reflection 
relation on Barnes' $G(1+N)$ function. Kinkelin's relation gives asymmetry 
of the $G(1+N)$ function (which is essentially the volume function for $SU(N)$ 
groups)  under $N\leftrightarrow -N$ transformation (which is  equivalent 
of the permutation of parameters, for $SU(N)$ groups), and coincides with 
universal relation on permutations at the $SU(N)$ line on Vogel's plane. 
These results are also applicable to universal partition function of 
Chern-Simons theory on three-dimensional sphere. 

This effect is  analogous to modular covariance, instead of invariance, of partition functions of appropriate gauge theories under modular transformation of couplings.}

\tableofcontents

\section{Introduction}

\subsection{ N $\leftrightarrow$ -N}

$N \leftrightarrow -N$ transformation is the symmetry of simple Lie 
algebras and gauge theories. E.g. dimensions of irreps of $SU(N)$ groups 
for a given Young diagram $Y$ can be represented as a
rational functions of $N$ and in this form it can be uniquely continued
 to an arbitrary $N$. After change of the sign of $N$ they are equal 
to dimensions of irreps with transposed Young diagram, up to the sign 
$(-1)^{Area(Y)}$. Similarly dimensions of $SO(N)$ transforms into those 
of $Sp(N)$, again with sign change and transposition of Young diagram 
\cite{King}. Eigenvalues of appropriate Casimir operators, 
analytically continued in a similar way, have the same 
behavior \cite{Cvitbook,MV11}. 

Parizi and Sourlas \cite{PaSo} noticed that space with $N$ odd grassmanian coordinates can be considered in some respects as a space with negative number $(-N)$ of usual even coordinates. This is in agreement of abovementioned symmetry w.r.t. the change of the sign of the $N$, since transposition of Young diagram interchange symmetrization and antisymmetrization. All this became a part of the theory of superalgebras and particularly  is formulated as an isomorphism of superalgebras $SU(n|m) \cong SU(m|n), OSp(n|m)\cong OSp(m|n)$. E.g., taking into account that many invariants in superalgebras depend on $n-m$, from $SU(n|m) \cong SU(m|n)$ we obtain $SU(n) \cong SU(-n)$. These dualities appear to be relevant in applications: $SU(N)$ gauge theory since first work of 't Hooft \cite{H1} is well-known to have $1/N$ expansion over even powers of $1/N$, $SO(N)$ gauge theories are dual to $Sp(N)$ theories \cite{Mkr}, with the same correspondence in representations of matter multiplets, and similarly in many other applications.

\subsection{Universality}
In a more recent time, after work of Vogel \cite{V0}, $N \leftrightarrow -N$  dualities became a part of invariance of theories under permutation of Vogel's parameters $\alpha, \beta, \gamma$, in the range of applicability of both notions. Note that ranges of applicability of universality and $N \leftrightarrow -N$ duality overlap, but neither is included in other one. E.g. universality, as of now, is dealing with adjoint and its descendant representations, while $N \leftrightarrow -N$, as described above, deals practically with all representations, but of classical groups, only. 

Vogel, motivated by knot theory, studied what 
can be called group weights 
of vacuum Feynman diagrams of gauge theories, 
but without any initially 
assigned Lie group. The problem he addressed was 
finally aimed to classify  
so called finite Vassiliev's invariants of knots, 
but during research 
he introduced very convenient parametrization 
of simple Lie algebras.  
These are so called universal, Vogel's, projective 
(i.e. relevant up to an arbitrary rescaling) parameters 
$\alpha, \beta, \gamma$ (see for details \cite{V0}).
They  can be defined as follows.

   Let $\gg$ be an arbitrary simple Lie algebra.
Consider symmetric square of its adjoint representation.
It can be canonically decomposed 
\cite{V0} into three irreducible representations:
\begin{equation*}
 S^2 ad=\underline{\textbf{1}}+Y_2(\alpha)+
  Y_2(\beta)+Y_2(\gamma)\,. 
\end{equation*}
Take a second Casimir operator $C_2$,
which is uniquely defined up to a scalar multiplier.
Denote by $2t$ eigenvalue of $C_2$ 
on the adjoint representation: $C_2(ad)=2t$. 
  Then the parameters $\a,\beta,\gamma$ are defined
through values of the Casimir $C_2$ on these irredicble
representations in the following way:
 \begin{equation}\label{defofvogelparameters}
 \begin{matrix}
C_2(Y_2(\alpha))=4t-2\alpha\,,  \cr  
C_2(Y_2(\beta))=4t-2\beta\,,   \cr  
C_2(Y_2(\gamma))=4t-2\gamma\,.   \cr  
\end{matrix}
\end{equation}
 One can show  that 
        \begin{equation}\label{defofvogelparameters1}
   \alpha+\beta+\gamma=t\,.
       \end{equation}
\noindent We see that these parameters 
are defined up to  a rescaling.
Permutation symmetry between them follows 
since there is no special 
order in these representations. 

\begin{definition}
Parameters $(\a,\b,\g)$  are called {\it Vogel's parameters}. 
 They can be considered as homogeneous coordinates
on projective plane ${\bf C} P^2$. Plane ${\bf C} P^2$ factorised
under the action of group $S_3$ of permutation of homogeneous
coordinates $\a,\b,\g$ is called {\it Vogel plane}.
\end{definition}

The values of Vogel parameters for all simple Lie algebras 
are given in table \ref{tab:1}, where for exceptional 
line $Exc(n), n=-2/3,0,1,2,4,8$ for  $G_2, D_4, F_4, E_6, E_7, E_8$
  respectively. (See \cite{Del,DM} for study of 
universality on exceptional line.)
\begin{table}[h] \label{tab:1}
\caption{Vogel's parameters for simple Lie algebras}     
\begin{tabular}{|r|r|r|r|r|} 
\hline Algebra/Parameters & $\alpha$ &$\beta$  &$\gamma$  & t \\ 
\hline sl(N) & -2 & 2 & N & N \\ 
\hline so(N) & -2  & 4 & N-4 & N-2 \\ 
\hline sp(N) & -2  & 1 & N/2+2 & N/2+1 \\ 
\hline Exc(n) & -2 & 2n+4  & n+4 & 3n+6 \\ 
\hline 
\end{tabular} 
\end{table} 
Parameter $\alpha$ is chosen to be equal to $-2$. 
This always can be done due to the scaling invariance. 
This choice ("minimal normalization") is distinguished by the 
fact that $t$ becomes an 
integer, the dual Coxeter number of corresponding algebra.  
The square of long roots in this normalization is equal to $2$.

\begin{example}\label{changingofsignexample}
Duality $N\leftrightarrow -N$ is implicitly present in table \ref{tab:1}  since Vogel's parameters are defined up to rescaling and permutation. 

Indeed, we see from the table
that  transformation $N\leftrightarrow -N$
 for $sl(N)$ is
reduced to the switching of parameters
$\a$ and $\beta$ and multiplication on $(-1)$:
$(-2,2,-N)=(-1)\cdot(2,-2,N)$.
In the same way under changing of sign of $N$
  $so(N)$ transforms into $sp(N)$ since
          $(-2,4,-N-4)=(-2)\cdot (1,-2,N/2+2)$.   
\end{example}


   Consider some quantity for simple Lie algebras, for example
dimension of algebra, dimensions of representations $Y_2(\,.\,)$,
eigenvalues of Casimir operators on irreducible representations, etc.
The "reasonable" function on Vogel plane, which for points
corresponding to simple Lie algebras 
(see table \eqref{tab:1}) takes the  values of that quantity on that 
Lie algebra, will be called   universal function 
corresponding to this quantity. 
For example, dimensions of simple Lie  algebras are 
given by the following universal dimension function

\begin{equation} \label{dim}
dim  =\frac{(\alpha-2t)(\beta-2t)(\gamma-2t)}{\alpha\beta\gamma}\,,
\quad (t=\a+\beta+\gamma)\,.
\end{equation}

Examples of universal functions include dimensions of some series 
of representations in powers of adjoint representation\cite{LM1}, 
eigenvalues of higher Casimir operators \cite{MSV}, 
characters of some representations on Weyl line  
and particularly such character for adjoint 
representation \cite{MV1,Westbury}:
\begin{eqnarray}\label{gene}
f(x)&=& \chi_{ad}(x\rho)=  
 r+\sum_{\mu} e^{x(\mu,\rho)} =\cr 
&=& \frac{\sinh(x\frac{\alpha-2t}{4})}{\sinh(x\frac{\alpha}{4})}
    \frac{\sinh(x\frac{\beta-2t}{4})}{\sinh(x\frac{\beta}{4})}
    \frac{\sinh(x\frac{\gamma-2t}{4})}{\sinh(x\frac{\gamma}{4})}\,,\quad
     (t=\a+\beta+\gamma)\,, 
\end{eqnarray}
where $r$ is the rank of simple Lie algebra, 
$\mu$ runs over the set of all roots of this algebra,
  and $\rho$ is the Weyl vector, which is equal to the half of the sum
of all positive roots:
       \begin{equation}\label{weylvector}
     \rho=\frac{1}{2}\sum_{\mu>0}\mu\,.
            \end{equation}
One can expect an existence of universal expression for quantities, 
related with adjoint representation, as those mentioned above. 
On the other hand,  there is no known universal expression  
for e.g vectorial representations of classical groups. 

We shall not discuss here the problem of what kind of 
analytic continuation on entire Vogel's plane in universal 
formulae is implied, since in this paper we actually deal with 
few explicitly defined analytic functions.

\subsection{Problem}
The aim of the present paper is to study the $N \leftrightarrow -N$ 
duality and Vogel's permutation symmetry of parameters in more 
complicated cases than the case when unversal function
is just a rational function of Vogel's parameters.  
The main objects will be the universal function of group's volume and universal partition functions of Chern-Simons theory on three dimensional sphere. 
Invariant volume of compact version of simple Lie group can 
be considered as partition function of corresponding 
matrix model \cite{Marino05}. Partition  function 
of Chern-Simons theory on three-dimensional sphere 
is calculated in terms of gauge group objects in \cite{W1}. 
Volume function also is the part of full Chern-Simons 
partition function. Both are represented in the
universal form in \cite{M13}. We shall see 
that corresponding analytic functions are not invariant 
w.r.t. the $ N \leftrightarrow -N$ duality  
and permutations of Vogel's parameters (as naively expected), 
but instead transform according to some 
non-trivial representations of these permutations' groups. 

For example, partition function of $SU(N)$ matrix model is 
essentially Barnes' $G$-function 
$G(1+N)$. 
At large $N$  its asymptotic expansion indeed is 
a series over $1/N^2$ (see e.g. \cite{Adamchik}), in agreement with  't Hooft 
perturbation theory observations \cite{H1} : 
\begin{eqnarray}
\log G(1+N)= \cr  \nonumber
\left(\frac{1}{2}N^2-\frac{1}{12}\right)\log(N)+
\frac{1}{2}N\log(2\pi)
 -\frac{3}{4}N^2+\frac{1}{12}-\log A+\cr  \nonumber
\sum_{g=2}^{\infty} \frac{B_{2g}}{2g(2g-2)}N^{2-2g}\,. 
\end{eqnarray}
where $B_{2g}$ are Bernoulli numbers. 

However, small $N$  expansion includes both even and 
odd powers \cite{Adamchik}:
\begin{eqnarray}
2\log G(1+N)=N \log (2\pi)-\gamma N^2 - N(N+1)+ \\ \nonumber
2\sum_{k=2}^{\infty}(-1)^k \zeta(k) \frac{N^{k+1}}{k+1}\,.
\end{eqnarray}

The $N \leftrightarrow -N $ asymmetry   is given by 
Kinkelin's relation \cite{Kin}: 
\begin{eqnarray} \label{Kinzero}
\log \frac{G(1+N)}{G(1-N)}= N\log(2\pi )- 
\int_{0}^{N}dx \, \pi x \, \cot (\pi x)\,.
\end{eqnarray}
So, the volume of $SU(N)$ is an analytical 
function $G(1+N)$, which is not invariant w.r.t. 
the $N \leftrightarrow -N$ duality. Functions $G(1+N)$ 
and $G(1-N)$ combine into doublet under duality transformation.

We are going to generalize these observations.
We will present the universal
formula for group's volume (and for Chern-Simons theory), 
will show that this universal formula
 defines few analytic functions,  and  
 will calculate  transformation of these functions
under permutations of Vogel's parameters. 

In more details: we will define this universal volume function 
by integral representation, which
 turns out to be
piecewise-analytical function of Vogel's parameters $\alpha,\beta,\gamma$.
Next we will show that this
integral representation  gives rise to six analytical functions on 
$CP^2$, constituting two triplet representations of group 
of permutations of Vogel's parameters. To study the 
behavior of these functions under permutation of arguments 
one has to analytically continue  them to the range of 
parameters larger than initially defined by integral 
representation. In this way we first calculate difference between functions from different triplets and then we calculate the difference between 
initial function and that with parameters permuted. 

As a check of this approach we specialize these results for the $SU(N)$ line on Vogel's plane. As mentioned above, in that case volume function is essentially Barnes' $G$-function, permutation of arguments is equivalent to changing of the sign of $N$, and relation between functions with permuted arguments leads to relation between $G(1\pm N)$, which will exactly coincide with Kinkelin's reflection relation (\ref{Kinzero}). 



\section{Universal invariant volume of simple Lie groups}
     Various formulae for volume of compact  simple Lie
groups are given by Macdonald \cite{Ma} (see also \cite{Hash}), 
Marinov \cite{Mar}, Kac and Peterson \cite{KP}, and Fegan \cite{Feg}. 
For few series of (super)groups volume formulae are 
given by Voronov \cite{Voron}. 
 In this section we derive the universal expression 
   for  volume of compact Lie groups, which  
generalizes these formulae for an arbitrary 
points on Vogel's plane,
and which presents them
 in a uniform way.
 
 This universal formula was obtained 
in \cite{M13} from the more general universal expression 
for perturbative part of Chern-Simons partition function.
  Universal volume formula is defined by a function
which coincides, at points from Vogel's table, 
with volume of corresponding groups: 
                    \begin{equation}\label{volume1}
          Vol(\a,\b,\g)=
           Vol\left({G}\left({\gg}_{[\a,\b,\g]}\right)\right)\,,
                    \end{equation}  
where $\gg=\gg_{(\a,\b,g)}$ is simple Lie algebra $\gg$ with coordinates
$(\a,\b,\g)$ on Vogel plane (see table \eqref{tab:1});  
  $G=G(\gg)$ is connected, simply connected compact Lie group
corresponding to Lie algebra $\gg$, and
$Vol(G)$ is  a volume of group's manifold with invariant metric. 
An invariant metric on group  is induced by certain invariant 
scalar product $(\,,\,)$ 
on the Lie algebra. On the simple Lie algebra an invariant scalar 
product  is proportional to Cartan-Killing form. On the other hand
 this scalar product defines canonically second Casimir $C_2$, which
in its turn defines Vogel's parameters $\a,\beta,\gamma$
by equations \eqref{defofvogelparameters},
\eqref{defofvogelparameters1}.  
 If $(\x,\y)=\lambda\phi(\x,\y)$,
where $\phi(\,,\,)$ is Cartan-Killing form, 
then $C_2$ has eigenvalue ${1\over \l}=2t$ on the Lie algebra.
Thus  Vogel's parameters $\a,\beta,\gamma$
of Lie algebra define invariant scalar product by equation
     \begin{equation}\label{scalingofmetric} 
    (\x,\y)={1\over 2t}\phi(\x,\y)\,,\quad (t=\a+\beta+\gamma)\,.
     \end{equation}
This is a scalar product which defines a metric 
of the group $G={G}\left({\gg}_{(\a,\b,\g)}\right)$ and its 
volume in equation\eqref{volume1}.
Under rescaling of Vogel's parameters volume changes in the following way:
         \begin{equation*}\label{scalingofvolume}
     V(\l\a,\l\beta,\l\gamma)=
 \l^{-{{\rm dim\,}}\over 2}V(\a,\beta,\gamma)\,.
         \end{equation*}

\subsection{Volume function and Chern-Simons partition function}

Recall briefly  construction of \cite{M13}, 
 obtained by considerations 
of  partition function of Chern-Simons theory. 
   
   Let $G$ be the compact Lie group. 
 In  \cite{MV1}  it was considered partition function
 $Z=Z^{(G)}(\k)$ for Chern-Simons theory corresponding to group
$G$ on $3$-dimensional sphere with coupling constant $\k$.
The partition function $Z^{(G)}(k)$ can be represented as
a product of perturbative and not-perturbative parts
$Z^{(G)}(\kappa)=Z^{(G)}_1 Z^{(G)}_2$, where non-perturbative part 
$Z^{(G)}_1$ is shown to be equal, 
on the basis of Macdonald formula \cite{Ma}, to   
             \begin{equation} \label{z1}
Z^{(G)}_1=\frac{(2\pi  \delta^{-1/2})^{dim }} {Vol(G)}\,,\quad
\delta=\kappa+t=\k+\a+\b+\g\,,
               \end{equation}
and perturbative part $Z^{(G)}_2$ is equal to
     \begin{equation}\label{z2}
Z_2=\prod_{\mu >0}
           { \frac{ 
\sin \frac{\pi (\mu,\rho)}{\delta} }
         {\frac{\pi (\mu,\rho)}{\delta}
         }}\,,
\end{equation}

 Here $\a,\beta,\gamma$ are Vogel's parameters of simple Lie algebra
which corresponds to group $G$. 

   $Vol(G)$ is the volume \eqref{volume1}
of the corresponding compact group $G$, 
product  $\prod_{\mu>0}$ is performed
over all positive roots $\mu$ of 
Lie algebra $\gg$, $\rho$ is
the Weyl vector \eqref{weylvector}.  
and $(\,,\,)$ is invariant scalar product\eqref{scalingofmetric}. 

Note that now scaling transformation of Vogel's parameters 
is extended to $\kappa$: $(\alpha, \beta, \gamma, \kappa) \rightarrow (\lambda \alpha,\lambda \beta, \lambda\gamma,\lambda \kappa)$, and Chern-Simons theory is invariant with respect to exactly this transformation.

   An important observation is that partition function
$Z(k)$ obeys  the condition
       \begin{equation}\label{partfunctionidentity}
     Z(\k)=1\,\, {\rm if}\,\, \k=0\,.
       \end{equation}
This immediately implies the volume formula (taking into account that  $Vol(G)$ doesn't depend on $\kappa$):

\begin{equation}\label{v1}
Vol(G)={(2\pi t^{-1/2})^{dim }\over Z_2^{(G)}} 
=(2\pi t^{-1/2})^{dim }   \prod_{\mu >0}
         \left(
{\sin \frac{\pi (\mu,\rho)}{t}/ \frac{\pi (\mu,\rho)}{t}}
          \right)\,.
\end{equation}

\begin{remark} {\it Chern-Simons partition function.}

Partition function $Z(\k)$ is equal to $S_{00}$, where  $S_{00}$ is 
the $(0,0)$ element of the matrix $S$ of modular 
transformations of characters of corresponding 
affine Kac-Moody algebra. Here $\kappa$ is coupling constant in front of the Chern-Simons action, which is rescaled 
simultaneously with Vogel's parameters and becomes 
integer (the level of representation) just in normalization 
of table \ref{tab:1}.
Since there is no  non-trivial representations at level zero 
the $S$ matrix becomes unit in that case.
  This implies condition
\eqref{partfunctionidentity}.  

\end{remark}

\subsection{Kac-Peterson formula.}

Kac and Peterson in 1984 derived an expression   for the 
volume of compact (connected, simply connected) 
simple Lie group defined with Cartan-Killing metric \cite{KP} 
(see also \cite{Feg}):
\begin{eqnarray}\label{KP}
Vol(G)=(2\sqrt {2} \pi)^{dim} \prod_{\mu >0} 
\frac{\sin 2\pi \phi(\rho, \mu)}{2\pi \phi(\rho, \mu)}\,,
\end{eqnarray}
where product is over positive roots of Lie algebra, 
and  $\phi(\,,\,)$ is Cartan-Killing form.

\begin{remark}\label{KPonly}
   Kac-Peterson formula can be immediately
deduced from our formula \eqref{v1} for volume.
Indeed according to \eqref{scalingofmetric}
scalar product  defining metric of the group
coincides with Cartan-Killing form if $t={1\over 2}$.
  If this condition is obeyed then
r.h.s. of equations \eqref{v1}
and \eqref{KP}  coincide.
    So, if one wishes, it is possible to completely 
discard Chern-Simons approach and start from equation \eqref{KP}, 
since equation \eqref{v1} is obviously equivalent to equation
\eqref{KP}. 
\end{remark}

\subsection{Universal expressions}

Now we rewrite the expressions for volume function and partition
function in universal form, following \cite{M13}.

It suffices to rewrite  universal expressions for \eqref{z2}, since
the universal formula 
for volume form \eqref{v1} can be expressed via this function due
to equation \eqref{dim}.
    
We have 
\begin{equation*}
\log Z^{(G)}_2=\sum_{\mu >0}\log \left( 
    \frac{\sin(\pi (\rho,\mu)/\delta)}{\pi (\rho,\mu)/\delta}
         \right)\,. 
    \end{equation*}
    \begin{equation}
=\sum_{\mu>0}
           \left(
       \log\left(\G\left(
    1-{(\rho,\mu)/\delta})
      \right)\right)+
         \log(\Gamma(1+(\rho,\mu)/\delta))
       \right)\,,
\end{equation}
where we use well-known representation
\begin{equation*}
\frac{\sin(\pi x)}{\pi x}=\frac{1}{\Gamma(1-x) \Gamma(1+x)}\,.
\end{equation*}
  Next  using the following
 integral representation of gamma-function:
  \begin{eqnarray}\nonumber
\log\Gamma(1+z)=\int_{0}^{\infty} 
 \frac{e^{-zx}+z(1-e^{-x})-1}{x(1-e^{-x})}dx\,,
\end{eqnarray}
we come to the  integral expression for perturbative partition function:
\begin{equation}
\log Z^{(G)}_2=
             -\int_0^\infty 
                        {
                 \sum_{\mu >0} 
                     \left(
   e^{x\frac{(\rho,\mu)}{\delta}}+e^{-x\frac{(\rho,\mu)}{\delta}}-2\,.
                       \right)
                       \over
              e^x-1
                  }dx
    \end{equation}
It follows from  equations \eqref{dim} and \eqref{gene} that:
      $$
\sum_{\mu >0} 
       \left(
   e^{x\frac{(\rho,\mu)}{\delta}}+e^{-x\frac{(\rho,\mu)}{\delta}}-2
       \right)
=f(x/\delta|\a,\beta,\gamma)- dim(\a,\beta,\gamma)\,.
      $$
where we stress dependence of $f(x)$ and $dim$ from universal parameters.

We use special notation \cite{MV1} for universal function in r.h.s.: 
       $$
F(x)=F(x|\a,\b,\g)= f(x|\a,\b,\g)-dim(\a,\b,\g)\,,
    $$
and arrive at universal formula
    \begin{equation} \label{z2u}
\log Z_2(\a,\beta,\gamma)=-\int^{\infty}_0  
\frac{ F(x/\delta)}{x(e^{x}-1)}dx\,.
   \end{equation}
In the same way as in equation \eqref{volume1}
we denote by $Z_2(\a,\b,\g)$ a function on
Vogel parameters $(\a,\beta,\gamma)$ such that
it coincides with the function $Z_2^{(G)}$ if $(\a,\beta,\gamma)$
are Vogel's parameters of Lie algebra $\gg$ corresponding to Lie group $G$.

Now using  
equation \eqref{v1}
for volume function we come to 
final universal expression for 
volume function $Vol(\a,\b,\g)$:
\begin{equation} \label{vol1}
 Vol(\a,\b,\g)= (2\pi t^{-1/2})^{dim}
                \exp{
          \left( -\int^{\infty}_0 
    \frac{F(x/t)}{(e^{x}-1)}{dx\over x}
             \right)\,,
                } 
            \end{equation}
and to universal expression for Chern-Simons partition function on $S^3$:
\beq \label{fullCS}
\log Z =  \int^{\infty}_0 
    \frac{F(x/t)-F(x/\delta)}{(e^{x}-1)}{dx\over x}= 
    \int^{\infty}_0  \frac{f(x/t)-f(x/\delta)}{(e^{x}-1)}{dx\over x}\,.
\eeq

We would like to emphasize again that
universal volume formula\eqref{vol1}  can be deduced
straightforwardly from Kac-Peterson formula \eqref{KP} 
disregarding all considerations
related with  Chern-Symon partition function
(see remark \ref{KPonly}).

\section{Volume function as Barnes' quadruple gamma-functions}

Our main aim is to study properties of analytical volume functions. 
It seems reasonable to establish connection of these functions with 
known functions such as Barnes' multiple gamma functions. 
In this section we will express volume function through Barnes' 
multiple gamma functions, following \cite{M13-2}.  We first 
recall definition of Barnes' multiple gamma
functions, then  we will formulate a proposition. Using this proposition  
 we express universal formulae \eqref{z2u} and 
\eqref{vol1} for perturbative partition function $Z_2$ and volume of group
in terms of Barnes's multiple gamma functions.

However, for further progress we need a developed theory of that 
functions as analytic functions of both argument and parameters, 
which is seemingly absent.

  Barnes' multiple ($N$-tuple) gamma function
 $\G_N=\G_N(w|a_1,\dots,a_N)$ can be defined 
 via Barnes' multiple zeta-function $\zeta_N=\zeta_N(s,w|a_1,\dots,a_N)$ 
in the following way \cite{Barnes1,Rui}:
         \begin{equation}\label{barnes1}
\G_N(w|a_1,\dots,a_N)=\exp
                     \left(
              {\p\over \p s}\z_N(s,w|a_1,\dots,a_N)\big\vert_{s=0}
                    \right)\,,
\end{equation}
where multiple zeta-function $\zeta_N(s,w|a_1,\dots,a_N)$  
  is a function on complex variables $s,w$ such that it is defined
for $\Re s>N$
 by power series
          \begin{equation}\label{zeta1} 
        \z_N(s,w|a_1,\dots,a_N)=\sum{1\over (w+k_1a_1+\dots+k_Na_N)^s}\,,
            \end{equation}
where summation goes over all non-negative integers $k_1,\dots,k_N$. 
It has meromorphic continuation in $s$ with simple poles
only at $s=1,2,\dots,N$.

Parameters $\{a_1,\dots  a_N\}$ are complex numbers which
obey the following condition: there exist a line passing through the origin,
such that all parameters are on the same side of this line.

   Barnes' zeta function obviously obeys the scaling condition:
   for every complex number $\l$,
                       \begin{equation}\label{scaling0}
        \z_N(s,\l w|\l a_1,\dots,\l a_N)= 
     \l^{-s}\z_N(s,w|a_1,\dots,a_N)\,,
                \end{equation}
and recurrent relations: 
                    \begin{equation}\label{translation0}
        \z_N(s, w+a_i| a_1,\dots, a_N)= \z_N(s,w|a_1,\dots,a_N)-
           \z_{N-1}(s,w|a_1,\dots,a_{i-1},a_{i+1},\dots,a_N)\,.
                \end{equation}
It is very useful to establish integral representation for
Barnes' function \eqref{barnes1}. We do it first 
for Barnes' zeta-function. We have 
          \begin{equation}\label{zetaintegral1}
        \z_N(s,w|a_1,\dots,a_N)={1\over \G(s)}\int_0^\infty x^{s-1}A(x)dx\,,
          \end{equation}
where  
         \begin{equation}\label{mainpart}
  A(x)={e^{-wx}\over\prod_{j=1}^N (1-e^{-a_j x})}\,.
        \end{equation}
Indeed, it is easy to see that r.h.s. of equations 
\eqref{zetaintegral1} and \eqref{zeta1} 
coincide for $\Re s >N$, by expansion of the integrand over powers of exponents. To calculate zeta-function for other $s$,
and in particular for $s=0$
we consider expansion of function $A(x)$ defined by equation
\eqref{mainpart}
in a vicinity of origin:  
               \begin{eqnarray}
 A(x)={e^{-wx}\over\prod_{j=1}^N (1-e^{-a_jx})}= 
           {1\over x^N\prod_{j=1}^N a_j}+\dots= \\
           =\sum_{k=-N}^{\infty}A_kx^k =
             \label{expansion1}
                 A_-(x)+A_0+A_+(x)\,, \\
\quad {\rm where\,\,} 
    A_-(x)=\sum_{k< 0}A_kx^k\,,\,\, A_+(x)=\sum_{k>0}A_kx^k\,,
              \end{eqnarray}
\begin{remark}\label{bernoul}
Coefficients of this expansion are {\it multiple Bernoulli polynomials} $B_n(w|a_1,\dots,a_N)$:
          \begin{eqnarray}
    A(x)={1\over x^N}\sum_{n=0}^{\infty}{(-1)^nx^n\over n!}
      B_{N,n}(w|a_1,\dots,a_N)
           \end{eqnarray}
          
          In particular
          \begin{eqnarray}
    A_-(x)=\sum_{n=0}^{N-1}{(-1)^nx^{n-N}\over n!}
  B_{N,n}(w|a_1,\dots,a_N)\,,
         \\
      \label{bernoulli0}
    A_0(x)={(-1)^N\over N!}B_{N,N}(w|a_1,\dots,a_N)\,.
         \end{eqnarray}
\end{remark}

   Let's perform meromorphic continuation in variable $s$
of multiple zeta-function.
  We use integral representation \eqref{zetaintegral1}.
It is well-defined, particularly, if
     \begin{equation}\label{condition0}
      \Re w>0, \Re a_i >0\,.
       \end{equation}
  Assume that condition \eqref{condition0} is obeyed.
 Then represent integral \eqref{zetaintegral1} as sum of integrals
from $0$ to $1$, and  from $1$ till infinity.
Integral $\int_1^\infty$ converges and 
it is an analytical function on $s$.
Using expansion \eqref{expansion1} , and the fact that 
meromorphic continuation
of the function $f(s)=\int_0^1 x^{s-1+n}dx$ is equal to
function ${1\over s+n}$,
we perform meromorphic continuation in $s$ of integral
$\int_0^1$.  Thus we perform meromorphic continuation
in $s$ of integral\eqref{zetaintegral1}.
 In particular for point $s=0$  
we come to the following
answers. Using expansions \eqref{expansion1} 
we see that for small $s$

\begin{eqnarray} \label{proofof}
\zeta(s,w|a_1,\dots,a_N)= 
                 {1\over \G(s)}\left(\sum_{k\leq 0} {A_k\over k+s}\right)+ \\
  {1\over \G(s)}\left(\int_0^1 x^{s-1}A_+(x)dx+
   \int_1^\infty x^{s-1}A(x)dx\right)\,. 
      \end{eqnarray}
 $\G(s)\approx {1\over s}$ in the vicinity of origin,
hence this equation implies that  
             \begin{equation}\label{zetaatorigin}
          \zeta(0,w|a_1,\dots,a_N)=A_0={(-1)^N\over N!}B_{N,N}(w)\,,
              \end{equation}
where   $B_{N,N}(w)$ is multiple Bernoulli polynomial
defined by \eqref{bernoulli0}. 

  Performing  further elementary calculations for equation
 \eqref{proofof} in the case when condition 
\eqref{condition0} is obeyed 
 we come to 
the following integral representation of Barnes's function \eqref{barnes1}:
           \begin{equation}\label{barnesintegral1}
    \G_N(w|a_1,\dots,a_n)=\exp\left(
           \int_0^\infty
           \left(
        A(x)-A_-(x)-A_0e^{-x}
           \right){dx\over x}
          \right)\,,
         \end{equation}
 where  function $A(x)$, $A_-(x)$ and $A_0$ are  defined by
equations \eqref{mainpart} and \eqref{expansion1}. 
If condition \eqref{condition0} is not obeyed one
has to use also relations \eqref{scaling0} and \eqref{translation0}.

\begin{remark} 
Integral representation \eqref{barnesintegral1} of Barnes' function    
appears in \cite{Rui}. Modern review of the theory of multiple Barnes' functions see in \cite{Spi}.
 \end{remark}

   It is instructive to write down equations for transformations
of Barnes gamma-functions under rescaling and under
shift of argument on parameter. They follow from 
definition \eqref{barnes1} of Barnes' functions
 and corresponding properties of zeta-function
(see equations \eqref{scaling0},\eqref{translation0}).
Scaling property \eqref{scaling0} implies that 
      \begin{equation}\label{scaling1}
 \G_N(\l w|\l a_1,\dots,\l a_N)=
 \l^{-c}\G_N(w|a_1,\dots,a_N)\,,
         \end{equation}
where $c=\zeta_N(0,w|a_1,\dots, a_N)=A_0=
 {(-1)^N\over N!}B_{N,N}(w|a_1,\dots,a_N)$.
 This formula can be also deduced straightforwardly
from integral representation \eqref{barnesintegral1}, 
with the use of Frullani's integral.

  Recurrent relation \eqref{translation0} implies that
          \begin{equation*}\label{translation1}
\G_N(w+a_i|a_1,\dots,a_N)={\G_N(w|a_1,\dots,a_N)
                      \over
               \G_{N-1}(w|a_1,\dots,a_{i-1},a_{i+1},\dots,a_N)\,.
                       }
          \end{equation*}

  One can say that $N$-tuple Barnes' function
  $\G(w|a_1,\dots,a_N)$ is completely defined by function  $A(x)$
in equation \eqref{mainpart}. We shall call $A(x)$ the main term in the integral representation of given multiple gamma function, or simply main term. 
  
More formally equation \eqref{barnesintegral1} defines a linear map, the
functional 
                 \begin{equation}\label{linearmap}
 A(x)\mapsto \int_0^\infty \left(A(x)-A_-(x)-A_0e^{-x}\right){dx\over x}
                 \end{equation}
on the linear space of functions which 
have finite Laurent series in a vicinity of origin
and decrease exponentially at infinity.
      In particular the function 
$A(x)$ in equation \eqref{mainpart}) belongs to this class  
(if $\Re w, \Re a_i>0$), and the value
of the functional on this function is equal to
$\log \G_N(w|a_1,\dots,a_n)$.  This simple observation implies
the following proposition
on properties of multiple gamma-functions.

\begin{proposition}\label{physicalfunction}

Let  
$\{A_p(x)=A_p(x,w_i|a^{(p)}_1,\dots,a^{(p)}_{N_p})\}$
$(p=1,\dots,k)$
be a finite set of functions of the form \eqref{mainpart}

  \begin{equation*}\label{mainpart1}
  A_p(x)={e^{-w_px}\over \prod_{j=1}^{N_p}\left (1-e^{-a^{(p)}_jx}\right)}\,.
        \end{equation*}

Consider the linear combination of these functions:
         $$
     G(x)=\sum_{p=1}^k l_pA_p(x)=\sum_{p=1}^k
   {l_pe^{-w_px}\over \prod_{j=1}^{N_i} (1-e^{-a^{(p)}_jx)}}\,.
          $$

    If  $G(x)/x$ is non-singular  at origin, then
          \begin{equation}
   \exp\left[\int_0^\infty G(x){dx\over x}\right]
                 =
    \prod_{p=1}^k \left(
        \G_{N_p}(w_p|a^{(p)}_1,\dots a^{(p)}_{N_i})\right)^{l_i}\,,
          \end{equation}

   In the special case when function  $G(x)$ vanishes, $G(x)\equiv 0$,
we have    
 \begin{equation*}
  \prod_{p=1}^M \left(\G_{N_p}(w_p|a_1,\dots,a_{N_p})\right)^{l_p}\equiv 1\,.              
  \end{equation*}

\end{proposition}

\begin{remark}
Proposition \ref{physicalfunction} provides the rigorous proof of all 
identities between multiple gamma functions used in \cite{M13-2,M14}. 
We expect that it is contained in some papers of past or current 
century, however we did not find an exact reference.

\end{remark}

   This proposition is very useful for analyzis of
   volume formula \eqref{vol1}. Using the fact that 
integrand in equation \eqref{vol1} is non-singular
function and Proposition \eqref{physicalfunction} 
  we obtain the following result: 
\begin{eqnarray}\label{vol}
Vol(G)&=& (2\pi t^{-1/2})^{dim}\exp{\left( -\int^{\infty}_0 \frac{dx}{x} \frac{F(x/t)}{(e^{x}-1)}\right) }   \\ 
&=& \left( \frac{4 \pi^2}{t} \right)^{\frac{dim}{2}} \frac{\Gamma_4(v_1) \Gamma_4(v_2) \Gamma_4(v_3) \Gamma_4(v_7)  }{\Gamma_4(v_4) \Gamma_4(v_5) \Gamma_4(v_6) \Gamma_4(v_8) }\left( \frac{t}{\pi} \right)^{\frac{dim}{2}} \\
   &=& \left( 4 \pi\right)^{\frac{dim}{2}} \frac{\Gamma_4(v_1) 
     \Gamma_4(v_2) \Gamma_4(v_3) \Gamma_4(v_7)  }{\Gamma_4(v_4) 
\Gamma_4(v_5) \Gamma_4(v_6) \Gamma_4(v_8) }\,,
\end{eqnarray}
where
\begin{eqnarray}\label{vs}
v_1&=& 2t-2\alpha\,\\
v_2&=& t+\gamma\,\\
v_3&=& t+\beta\,\\
v_4&=& 3t\,\\
v_5&=& 2t+2\beta+\gamma,\\
v_6&=& 2t+\beta+2\gamma,\\
v_7&=& 5t-\alpha\,    \\ \label{vs2}
v_8&=& -\alpha\,.
\end{eqnarray}
parameters of functions $\Gamma_4$ are $(-\alpha,\beta,\gamma,2t)$, 
and we use equation \cite{Rui}:
\begin{eqnarray} 
\Gamma_1(x|x)=\sqrt{\frac{x}{2\pi}}\,. 
\end{eqnarray}

The scaling properties of volume functions now can be 
deduced from scaling properties of quadruple gamma functions. 
Direct calculation confirms that it is in agreement with \eqref{vol1}.

Finally, we introduce multiple sine functions \cite{Nar}
\begin{equation}\label{Sr}
S_r(w|a_1,a_2,...)=
\frac{\Gamma_r(|a|-w|a_1,a_2,...)^{(-1)^r}}{\Gamma_r(w|a_1,a_2,...)}\,,
\quad (|a|=\sum_{j=1}^{r}a_j)\,.
\end{equation}
Important feature of multiple sine functions is scaling invariance:
\begin{eqnarray}\label{scalinginvariance}
S_r(\l w|\l a_1, \l a_2,...)=S_r(w|a_1,a_2,...)\,.
\end{eqnarray}
Indeed, using linear map \eqref{linearmap} one can define 
\eqref{barnesintegral1}-like integral representation
of multiple sine function (see below equation \eqref{sineintegral}). 
It follows from definition \eqref{Sr} 
of multiple sine function and integral representation
\eqref{barnesintegral1} for Barnes's functions that
coefficient $A_0$ for multiple sine function vanishes. Thus
equation \eqref{scaling1} implies equation \eqref{scalinginvariance}.

 The reasonable question is whether simple scaling 
properties of volume function and its expression in terms of 
multiple gamma functions  lead to representation of volume in 
terms of multiple sine functions and some simple 
functions with necessary (non-trivial) scaling dimension.

With that purpose and with the help of Proposition \ref{physicalfunction}  
we transform volume function into
\begin{eqnarray}
Vol(G)=\left( 4 \pi\right)^{\frac{dim}{2}} \times  \\  \nonumber
\frac{S_4(\alpha+\beta+\gamma|-\alpha,\beta,\gamma,2t)S_4(2\beta+\gamma)S_4(\beta+2\gamma)}{S_4(2\alpha+3\beta+3\gamma)}
 \times  \\ \nonumber
S_3(-\alpha|-\alpha,\beta,\gamma) \times \\ \nonumber
\frac{\Gamma_2(\alpha+3\beta+3\gamma|\beta,\gamma)\Gamma_2(2\alpha+3\beta+3\gamma|\beta,\gamma)}{\Gamma_2(\beta+\gamma|\beta,\gamma)\Gamma_2(\alpha+\beta+\gamma|\beta,\gamma)}
 \times \\ \nonumber
\frac{\Gamma_2(-\alpha|-\alpha,\gamma)
\Gamma_2(-\alpha|-\alpha,\beta)\Gamma_1(2\beta+2\gamma|-\alpha)
\Gamma_1(\beta+\gamma|-\alpha)}{\Gamma_1(-\alpha|-\alpha)
\Gamma_2(2\beta+2\gamma|-\alpha,\beta)
\Gamma_2(2\beta+2\gamma|-\alpha,\gamma)}\,,
\end{eqnarray}
where all functions $S_4$ have the same parameters 
$(-\alpha,\beta,\gamma,2t) $. 

One can see that already double gamma functions 
 don't combine into double sine functions. 

Another important feature  of multiple sines is that their
\eqref{barnesintegral1}-like integral representation, 
can be transformed into the integral over entire real axis  \cite{Nar}:
\begin{eqnarray} \label{sineintegral}
& & \log S_r(z|\underline{\omega})=\\ \nonumber
&=& (-1)^r \frac{\pi i}{r!}B_{rr}(z|\underline{\omega})+(-1)^r \int_{R+i0}\frac{dx}{x}\frac{e^{zx}}{\prod_{k=1}^r(e^{\omega_i x}-1)} \\ \nonumber
&=&(-1)^{r-1} \frac{\pi i}{r!}B_{rr}(z|\underline{\omega})+(-1)^r \int_{R-i0}\frac{dx}{x}\frac{e^{zx}}{\prod_{k=1}^r(e^{\omega_i x}-1)} 
\end{eqnarray}
We shall see below, that
integral representation in terms of integral over entire $x$ axis 
leads to "better" analytic properties of sine function with 
respect to its parameters. More exactly, in this case
zero is not 
the branch point of sine function as an analytic function 
of any of its parameters. We see above that 
volume function can't be represented as a product/ratio of multiple 
sine functions. However, full partition 
function of Chern-Simons theory can be expressed
via multiple sine functions, as shown in \cite{M14,KM}.

\section{Analytic functions from Chern-Simons perturbative partition function and 
    anomaly of Vogel's symmetry} \label{reflect}

Denote by $K(\alpha,\beta,\gamma)$
 the integral which appears in exponents  in expressions above   
$K(\alpha,\beta,\gamma)$
\begin{eqnarray}\label{K}
K(\alpha,\beta,\gamma)&=& \int^{\infty}_0 \frac{dx}{x} 
\frac{F(x)}{(e^{x}-1)}\,. \label{Kvol}
\end{eqnarray}
This reproduces the integral in equation (\ref{z2u}) for $\log Z_2$ , 
if all arguments are divided on $\delta$, or that 
for volume function (\ref{vol1}), if all arguments are divided on $t$.
Consider  $K(\alpha,\beta,\gamma)$ for an arbitrary
complex values of all variables, $\a,\beta,\gamma$.
  We put $\delta=1$ 
by scaling transformation. Next, it is evident, that one 
should have $\Re\alpha\neq 0, \Re\beta\neq 0, \Re\gamma\neq 0$ 
since otherwise there are non-integrable singularities at the poles 
of one of $\sinh$ in denominator in $F(x)$. This restriction divides the 
space of parameters into disjoint regions. Inside that 
regions integral can converge or diverge at large $x$ depending 
on values of parameters. It converges in all regions if values 
of moduli of $\delta$ are sufficiently 
large. In our  normalization ($\delta=1$ after rescaling), 
this means sufficiently small values of 
moduli of $\alpha, \beta, \gamma$.

Function $K$ is invariant w.r.t. the permutations of parameters, 
but in general it is not an analytic function of parameters. 
To understand what happens it is instructive
to consider the following toy model suggested in \cite{M13}: 
Consider the function $f$  such that 
\begin{equation}\label{toy1}
      f(z)=
\int_{0}^{\infty}\frac{dx}{\cosh(zx)}= 
           \begin{cases}
 \frac{\pi}{2z} &\mbox{if $\Re z >0$} \cr
 -\frac{\pi}{2z}&\mbox{if $\Re z <0$}  \cr
       \end{cases}\,,
\end{equation}
or one can consider even simpler integral representation
for this function (see \cite{M13-2}):
      \begin{equation}\label{toy2}
     f(z)=
\int_{0}^{\infty}\frac{dx}{1+(zx)^2}  =   
   \begin{cases}
 \frac{\pi}{2z} &\mbox{if $\Re z >0$} \cr
 -\frac{\pi}{2z}&\mbox{if $\Re z <0$} \cr
       \end{cases}\,.
              \end{equation}
Both integrals define  analytic function 
$f_+(z)=\pi/2z$ for positive real part of 
argument, 
 and  analytic function 
$f_-(z)=-\pi/2z$
for negative real part of argument.
On the other hand a function $f(z)$ is not anaylitic 
on the whole plane.
  The reason is that one cannot connect  two points on a 
complex plane of parameter $z$, one with positive real part and 
another one with negative, by a continuous path without passing through 
singularity of integrals (\ref{toy2})(\ref{toy1}). Namely, 
one can't avoid crossing the line $\Re z=0$, any point on which 
is singular for integrals. Values of integrals at $\Re z>0$ and  
$\Re z<0$ do not  belong to the same analytic function. 
  They are given instead by two different analytic functions 
$f_+(z)$ and $f_-(z)$ (for $z\not=0$).
   Each of these functions is initially defined in the corresponding 
region of convergence of integral (\ref{toy2}) or (\ref{toy1}), 
i.e. corresponding open half-plane.

  One can  take each of these functions, for example
  a function $f_+(z)$, 
  continue it analytically to the half-plane ${\rm Re\,}z<0$, then
compare the analytical continuation of function $f_+(z)$
with another function, $f_-(z)$. 
We see that they are 
related  by the transformation $z \rightarrow -z$, and 
their difference is $f_+(z)-f_-(z)=\pi/z$. This  is 
the simplest example  of reflection relation.

Let's introduce notation $K_{\pm\pm\pm}(\alpha,\beta,\gamma)$ for 
analytic functions which are equal to function 
$K(\alpha, \beta, \gamma)$ (defined by euation \eqref{K})
 in the regions,  where 
signs of real parts of 
 parameters coincide respectively with its indices.
\begin{eqnarray}\label{Kpm}
K_{\epsilon_1 \epsilon_2 \epsilon_3}(\alpha,\beta,\gamma)= 
\int^{\infty}_0 \frac{dx}{x} \frac{F(x)}{(e^{x}-1)}\,, \\
       \epsilon_1= sign (\Re \alpha), 
       \epsilon_2= sign (\Re \beta), 
       \epsilon_3=sign (\Re \gamma)\,.\nonumber
\end{eqnarray}
Functions $K_{\pm\pm\pm}$ are symmetric w.r.t. the 
transposition of arguments corresponding to the same signs in index, 
since we can interchange them smoothly by paths in the region 
of definition of integral. For example 
\begin{eqnarray}\label{symm}
K_{--+}(\alpha,\beta,\gamma)=K_{--+}(\beta,\alpha,\gamma)\\
\Re \alpha<0, \Re\beta<0, \Re\gamma>0   \nonumber
\end{eqnarray}
but in general it is not symmetric w.r.t. the transposition of $ \beta, \gamma$. 
From definitions we get relations:

\begin{eqnarray}\label{Kpmm}
K_{--+}(\alpha,\beta,\gamma)=K_{-+-}(\alpha,\gamma,\beta)=K_{+--}(\gamma,\alpha,\beta)\\
\Re \alpha<0, \Re\beta<0, \Re\gamma>0  \nonumber
\end{eqnarray}
and 
\begin{eqnarray}\label{Kppm}
K_{++-}(\alpha,\beta,\gamma)=K_{+-+}(\alpha,\gamma,\beta)=K_{-++}(\gamma,\alpha,\beta)\\
\Re \alpha>0, \Re\beta>0, \Re\gamma<0  \nonumber
\end{eqnarray}

   We shall analytically continue these functions to other regions, 
where they don't necessarily coincide with functions $K$ originated 
from that region, and would like to calculate their difference.  
So, for example, we take $K_{--+}(\alpha,\beta,\gamma)$, 
where $\Re \alpha<0, \Re\beta<0, \Re\gamma>0$, analytically continue 
it to other region of arguments, e.g. 
$\Re \alpha>0, \Re\beta<0, \Re\gamma>0$ and explicitly calculate 
difference  
$K_{--+}(\alpha,\beta,\gamma) - K_{+-+}(\alpha,\beta,\gamma)$. 
Carrying on this analytic continuation twice, w.r.t. the arguments 
with different signs of real part, and applying corresponding relations, 
we obtain the behavior of initial function $K$ under transposition of parameters with different signs of real part. 

Relations of type (\ref{symm}),(\ref{Kpmm}) and (\ref{Kppm}) will be maintained by these analytic continuations. 
 
So, let's consider some $K$ in the region of convergence of its parameters, 
and analytically continue one  parameter (say $\alpha$) 
with $\Re \alpha>0$ into region with  negative $\Re \alpha$, assuming 
integral  is still convergent. Integrand is regular function of $\alpha$ 
except the poles at the points where $\sinh(x\alpha/4)$ becomes zero, 
$x=0$ excluded, i.e. $x=4\pi i k/\alpha, k=\pm 1, \pm 2, ...$. 
Since $\Re \alpha >0$, poles with $k>0$ are located in upper half-plane.  
When we change $\alpha$, keeping module non-zero and changing  
argument in counterclockwise direction, poles move in clockwise direction. 
The integral remains convergent until half of poles (with $k>0$) 
reach integration contour $0 \leq x < \infty $, i.e. 
when some poles become real and positive.  When poles reach 
integration contour from upper half-plane and continue to move to 
lower half-plane, we deform contour to prevent appearance of singularity. 
One can imagine that deformation as a creation of a narrow sprout 
of the contour, which goes from the real positive line 
to a pole (which is in the lower half-plane), turns around it in 
counterclockwise direction, and return to real positive line. 
Moving parameter to its new value, and simultaneously deforming the contour, 
we get a value of initial function, analytically continued to new 
value of parameter. Then we substitute new contour by equivalent one, 
which consists from infinite number of pieces: 
one piece is again real positive line from zero to positive infinity, 
others are small counterclockwise circles around poles in the 
lower half-plane. Line integral is $K$ from new values of parameters, 
which is the $K$, originated from the region of new values of parameters,
 so the sum over ($ 2\pi i$ times residues of) poles gives 
difference between analytically continued function and that  $K$. 

    Let's write down all this for specific $K$, 
say $K_{+-+}(\alpha,\beta,\gamma)$. Initially it is 
defined for $\Re \alpha>0, \Re \beta<0, \Re\gamma>0$. 
We would like to analytically continue it on the region 
$\Re \alpha<0, \Re\beta<0, \Re\gamma>0$. 
According to above, we get:
\begin{eqnarray}\label{Reflect1}
K_{+-+}(\alpha,\beta,\gamma)=K_{--+}(\alpha,\beta,\gamma)+
\varphi_+(\alpha|\beta,\gamma),\\ \nonumber
\Re \alpha<0, \Re\beta<0, \Re\gamma>0\,, 
\end{eqnarray}
where
\begin{eqnarray} \label{phiplyus}
\varphi_+(\alpha|\beta,\gamma)=2\pi i \sum_{k=1}^{\infty} \frac{i e^{-\frac{2k i \pi }{\alpha}}  \sin \left[\frac{k \pi  (\beta-2 t)}{\alpha}\right] \sin\left[\frac{k \pi  (\gamma-2 t)}{\alpha}\right]}{2k \pi \sin\left[\frac{k \beta \pi }{\alpha}\right] \sin\left[\frac{k \gamma \pi }{\alpha}\right]} \\ \nonumber
=-\sum_{k=1}^{\infty} \frac{ e^{-\frac{2k i \pi }{\alpha}}  \sin \left[\frac{k \pi  (\beta+2\gamma)}{\alpha}\right] \sin\left[\frac{k \pi  (\gamma+2\beta )}{\alpha}\right]}{k  \sin\left[\frac{k \beta \pi }{\alpha}\right] \sin\left[\frac{k \gamma \pi }{\alpha}\right]}
\end{eqnarray}

This can be further transformed into

\begin{eqnarray}
\varphi_+(\alpha|\beta,\gamma)= \\ \nonumber
-\sum_{k=1}^{\infty}  \left(  \frac{ 2e^{-\frac{2k i \pi }{\alpha}}  \sin \left[\frac{2k \pi  \beta}{\alpha}\right] \cos\left[\frac{k \pi  \gamma}{\alpha}\right]}{k  \sin\left[\frac{k \gamma \pi }{\alpha}\right]}+  \frac{ 2e^{-\frac{2k i \pi }{\alpha}}  \sin \left[\frac{2k \pi  \gamma}{\alpha}\right] \cos\left[\frac{k \pi  \beta}{\alpha}\right]}{k  \sin\left[\frac{k \beta \pi }{\alpha}\right]} + \right. \\ \nonumber
\left. \frac{e^{-\frac{2k i \pi }{\alpha}}}{k} \left( 1+\cos \left[ \frac{2k \pi  \beta}{\alpha} \right] +\cos \left[ \frac{2k \pi  \gamma}{\alpha} \right] +2\cos \left[ \frac{2k \pi  \beta}{\alpha} + \frac{2k \pi  \gamma}{\alpha} \right] \right) \right) =\\ \nonumber
\log \left( \left(1-e^{2 i a}\right) \sqrt{1-e^{2 i (a-x)}} \sqrt{1-e^{2 i (a+x)}} \right. \\ \nonumber
\left.  \sqrt{1-e^{2 i (a-y)}} \sqrt{1-e^{2 i (a+y)}} \left(1-e^{2 i (a-x-y)}\right) \left(-1+e^{2 i (a+x+y)}\right) \right)- \\ \nonumber
-\sum_{k=1}^{\infty}  \left(  \frac{2 e^{2kia}  \sin [2kx] \cos\left[ ky\right]}{k  \sin\left[ky\right]}+  \frac{ 2e^{2kia}  \sin \left[2ky\right] \cos\left[kx\right]}{k  \sin\left[kx\right]} \right)\\
a=-\frac{\pi}{\alpha}, x=\frac{\pi \beta}{\alpha}, y=\frac{\pi \gamma}{\alpha}
\end{eqnarray}

Would we consider movement of parameter $\beta$ with initially negative $\Re \beta$, difference will appear in that contour of integration (positive line) will be reached by poles $x=4\pi i k/\beta$ with negative $k$, $k=-1,-2,...$. So in that case we get:  

\begin{eqnarray}\label{Reflect2}
K_{--+}(\alpha,\beta,\gamma)=K_{-++}(\alpha,\beta,\gamma)+ \varphi_-(\beta|\alpha,\gamma), \\ \label{phiminus}
\varphi_-(\beta|\alpha,\gamma)= -2\pi i \sum_{k=1}^{\infty} \frac{i e^{\frac{2k i \pi }{\beta}}  \sin \left[\frac{k \pi  (\alpha-2 t)}{\beta}\right] \sin\left[\frac{k \pi  (\gamma-2 t)}{\beta}\right]}{2k \pi \sin\left[\frac{k \alpha \pi }{\beta}\right] \sin\left[\frac{k \gamma \pi }{\beta}\right]} \\ \nonumber
\Re \alpha<0, \Re\beta>0, \Re\gamma>0 
\end{eqnarray}

Note properties of $\varphi_{\pm}(\alpha|\beta,\gamma)$, which follow from their definitions:
\begin{eqnarray}
\varphi_{\pm}(\alpha|\beta,\gamma)=\varphi_{\pm}(\alpha|\gamma,\beta),\\
\varphi_{\pm}(\alpha|-\beta,-\gamma)=\varphi_{\pm}(\alpha|\beta,\gamma),\\
\varphi_{-}(\alpha|\beta,\gamma)=-\varphi_{+}(-\alpha|\beta,\gamma)
\end{eqnarray}

It is easy to check that due to these properties relations \eqref{Kpmm}, \eqref{Kppm} are maintained after analytic continuations. 

\begin{remark}
Remaining sums over $k$ in functions $\varphi_{\pm}$ \eqref{phiplyus},\eqref{phiminus} can be further transformed due to following identity given in Appendix A1 of \cite{KM}, based on Jonqui\`ere's inversion formula for polylogarithm functions:  
\begin{eqnarray}
\sum_{k=1}^{\infty}\frac{e^{kA}}{k \sin kB} \approx -\sum_{k=1}^{\infty}\frac{e^{-kA}}{k \sin kB}
\end{eqnarray}
up to the simple terms, bilinear over  Bernoulli polynomials $B_0, B_1, B_2$.  We shall not do that, since it doesn't simplify expressions strongly enough. See, however, remark below.

\end{remark}
To obtain the change of functions $K$  under permutation of parameters, one have to extend in (\ref{Reflect1})  $\beta$ from the region $\Re \beta <0$ to  $\Re \beta >0$, i.e. apply (\ref{Reflect2}), using explicit form of functions $\varphi$ at all values of parameters:
\begin{eqnarray}\label{permpar0}
K_{+-+}(\alpha,\beta,\gamma)=K_{--+}(\alpha,\beta,\gamma)+\varphi_+(\alpha|\beta,\gamma)=\\
K_{-++}(\alpha,\beta,\gamma)+\varphi_+(\alpha|\beta,\gamma)+\varphi_-(\beta|\alpha,\gamma)=\\
K_{+-+}(\beta,\alpha,\gamma)+\varphi_+(\alpha|\beta,\gamma)+\varphi_-(\beta|\alpha,\gamma)\\
\Re \alpha<0, \Re\beta>0, \Re\gamma>0\,, 
\end{eqnarray}
where for the last equality we use (\ref{Kppm}): 
$K_{-++}(\alpha,\beta,\gamma)=K_{+-+}(\beta,\alpha,\gamma)$ 
at $\Re \alpha<0, \Re\beta>0, \Re\gamma>0  $. 
So, the  change of function $K$ under 
transposition of two arguments with 
different signs is given by equation
\begin{eqnarray}\label{permpar}
K_{+-+}(\alpha,\beta,\gamma)- K_{+-+}(\beta,\alpha,\gamma)=\varphi_+(\alpha|\beta,\gamma)+\varphi_-(\beta|\alpha,\gamma)\\ \nonumber
\Re \alpha<0, \Re\beta>0, \Re\gamma>0 \,.
\end{eqnarray}

 In the next section we shall show that on $sl(N)$ line this formula recovers Kinkelin's relation on Barnes' $G$-function.

\begin{remark}
In this section we consider some properties of volume 
function(s) as  an analytic  functions of parameters. This is necessary for 
calculation of result of permutation of volume function arguments at an arbitrary point of Vogel's plane. However, a lot of questions remain untouched. E.g. one can ask on an analytic continuation along the circular path around an origin. It is easy to show that for multiple sine functions, due to integral representation \eqref{sineintegral} with integration on entire $x$ axis, the similar (to above) deformation of contour of integration leads to the same value of function, i.e. zero is not a branch point of parameter(s). This remark is relevant for full Chern-Simons partition function on three dimensional  sphere, since it is expressed purely in terms of multiple sine functions \cite{M14,KM}. For a general multiple gamma functions one will obtain finite bilinear combination of Bernoulli polynomials, due to the abovementioned identities in  \cite{KM}. All that require separate study.  
\end{remark}

\section{Volume analytic functions for SU(N) and Kinkelin's reflection relation for Barnes' G-function}\label{reflectsun}

For the case $\delta=t (=1)$, it is easy to establish that  integral converges when parameters  $\Re \alpha, \Re \beta, \Re \gamma, (\alpha+\beta+\gamma=1)$ are of different signs, and diverges otherwise (i.e. when they all are positive). On the plane $(\Re\alpha, \Re\beta)$ line $\Re\gamma=0$ corresponds to line $\Re\alpha+\Re\beta=1$. So, lines of zero real parts of parameters divide  $(\Re\alpha, \Re\beta)$ plane on 7 regions. Similarly hyperplanes  $\Re\alpha=0, \Re\beta=0$ and  $\Re\gamma=0$ divide projective space of $\alpha, \beta, \gamma$ (i.e. $CP^2$) into seven disconnected pieces. It is easy to deduce that integral doesn't converge in one region  only, namely in the region where all real parts of parameters are positive.

Next we would like to make contact with Kinkelin's functional 
equation \cite{Kin} for Barnes' $G$-function 
(which is essentially Barnes' double gamma-function). 
For that purpose we shall apply our equation (\ref{permpar}) 
to the case when volume function is expressed via $G$-function, 
which happens for groups $SU(N)$.  Group
 $SU(N)$ corresponds  to parameters 
$(\alpha, \beta, \gamma)=(-2,2,N), \delta=t=N$. We remove 
constraint $t=1$, and explicitly leave $t=N$ in equations, 
for easier comparison with known results. Besides that, 
since some contributions are singular at these values, 
we take $\alpha=-2, \beta=2+x, \gamma=N-x, t=N $, 
and take a limit $x\rightarrow 0$. 
Then we have 
\begin{eqnarray}
\varphi_+(\alpha|\beta,\gamma)=2\pi i \sum_{k=1}^{\infty} 
\frac{i e^{-\frac{2k i t \pi }{\alpha}}  
\sin \left[ \frac{k \pi  (\beta-2 t)}{\alpha}\right] 
\sin\left[\frac{k \pi  (\gamma-2 t)}{\alpha}\right]}{2k \pi \sin\left[\frac{k \beta \pi }{\alpha}\right] \sin\left[\frac{k \gamma \pi }{\alpha}\right]}=\\ \nonumber
2\pi i \sum_{k=1}^{\infty}  
\left( \frac{-1+e^{2k i \pi  t}}{2k^2 \pi ^2 x}+\frac{i 
\left(1+4 e^{k i \pi  t}+e^{2k i \pi  t}\right)}{4k \pi }\right) +O[x]\,,
\end{eqnarray}

\begin{eqnarray}
\varphi_-(\beta|\alpha,\gamma)=-2\pi i \sum_{k=1}^{\infty} \frac{i e^{\frac{2k i \pi }{\beta}}  \sin \left[\frac{k \pi  (\alpha-2 t)}{\beta}\right] \sin\left[\frac{k \pi  (\gamma-2 t)}{\beta}\right]}{2k \pi \sin\left[\frac{k \alpha \pi }{\beta}\right] \sin\left[\frac{k \gamma \pi }{\beta}\right]}=\\   
2\pi i \sum_{k=1}^{\infty}
\left(  -\frac{-1+e^{2k i \pi  t}}{2k^2 \pi ^2 x} + \right.  \\ \left.  +
\frac{1-k i \pi -4k i e^{k i \pi  t} \pi +i e^{2k i \pi  t} 
(i+k \pi  (-1+2 t))}{4k^2 \pi ^2}\right) +O[x]\,. 
\end{eqnarray}
The sum is regular at $x\rightarrow 0$. It is equal to 
\begin{eqnarray}\label{sumregular}
2\pi i \sum_{k=1}^{\infty}\frac{1+e^{2k i \pi  N} 
(-1+2k i \pi  N)}{4k^2 \pi ^2}
\end{eqnarray}

So, from equation \eqref{permpar} we have that 
\begin{eqnarray} \label{permSU}
K_{-++}\left(
-\frac{2}{N},\frac{2}{N},1\right)-
K_{-++}\left(\frac{2}{N},-\frac{2}{N},1\right)=
\\  \nonumber
-2\pi i \sum_{k=1}^{\infty}\frac{1+e^{2k i \pi  N} 
(-1+2k i \pi  N)}{4k^2 \pi ^2}\,.
\end{eqnarray}

The same expression appears, when we calculate in reverse order:
  first put $SU(N)$ parameters into integral (\ref{Kpm}),
  and then calculate its asymmetry under transposition 
of parameters. Integral for $SU(N)$ is \cite{M13, M13-2} is equal to
\begin{equation}\label{suvol}
K_{-++}\left(-\frac{2}{N},\frac{2}{N},1\right)=
\int^{\infty}_0 \left(
\frac{1-e^{-x}}{4\sinh^2(\frac{x}{2N})}-\frac{N^2}{e^x-1}
                 \right)
              \frac{dx}{x}\,,\quad (\Re N > 0)\,.
\end{equation}
   For $SU(N)$, switching of parameters $\alpha, \beta$ is equivalent to transformation $N \rightarrow -N$. So we need a change of (\ref{suvol}) under the change of sign of $N$.  It can be done in the same way as above. Let we have $N$ with $\Re N >0$.  $N$-dependent poles of integrand of \ref{suvol} are in the points $x= \pm i \pi k/N, k= 1,  2, ...$. Now let's move $N$ to $-N$, e.g. by multiplying on phase factor, changing from 1 to -1 in counterclockwise direction.  Then poles will move in clockwise direction and those with $k>0$ will touch the integration line $[0,\infty)$.  To avoid singularity, we change contour as above. Finally, when $N$ becomes $-N$, we get new contour of integration and replace that by half-line from 0 to infinity and a small circles, enclosing poles at points  $x=- i \pi k/N, k= 1,  2, ...$ in counterclockwise direction.  Integral over half-line is an initial integral with $-N$ instead of $N$, which is the same. So, the value of analytically continued function at the  point $-N$ is equal 
to its value at the point $N$ plus $2\pi i$ times residues at poles. 
Residue in the pole at $x= - i \pi k/N$ is:

\begin{eqnarray}
Res_{x= -\frac{i \pi k}{N}} \left( \frac{1}{x}\left(\frac{1-e^{-x}}{4\sinh^2(\frac{x}{2N})}-\frac{N^2}{e^x-1}\right) \right)=\\ 
\frac{1+e^{2k i \pi  N} (-1+2k i \pi  N)}{4k^2 \pi ^2}\,.
\end{eqnarray}
So the sum coincides with  expression \eqref{sumregular}.

Now let's use this answer with integral representation \cite{M13} of Barnes' $G$-function \cite{Barnes2} in terms of integral \eqref{suvol}.
\begin{eqnarray} \label{GN}
\log (G(1+N))=\frac{1}{2}N^2\log N -\frac{1}{2}(N^2-N)\log(2\pi)+ \\ \nonumber
+K_{-++}(-\frac{2}{N},\frac{2}{N},1)\,.
\end{eqnarray}
From this equation, applying the procedure of sign changing 
of $N$ by counterclockwise rotation, and using (\ref{permSU}), 
we get reflection relation for Barnes' $G$-function:

\begin{eqnarray} \label{GNGN}
\log \frac{G(1+N)}{G(1-N)}=\frac{i\pi}{2}N^2 + \\  \nonumber
N\log(2\pi)- i \sum_{k=1}^{\infty}\frac{1+e^{2k i \pi  N} (-1+2k i \pi  N)}{2k^2 \pi}
\end{eqnarray}
provided we choose appropriate branch of $\log N$.

We would like to compare this with  Kinkelin's 
functional equation \cite{Kin},

\begin{eqnarray} \label{Kin0}
\log \frac{G(1+N)}{G(1-N)}= N\log(2\pi )- \int_{0}^{N}dx \, \pi x \, cot(\pi x)
\end{eqnarray}
in a form given in \cite{Adamchik}:
\begin{eqnarray} \label{Kin}
\log \frac{G(1+N)}{G(1-N)}=  \frac{i}{2\pi}Li_2(e^{2\pi i N})
+N\log\left( \frac{\pi}{\sin\pi N}\right) -\frac{\pi i}{2}B_2(N)\,,
\end{eqnarray}
where $Li_2$ is the dilogarithm function, $B_2(z)=z^2-z+1/6$ is 
second Bernoulli polynomial. These two forms of Kinkelin's relation 
are equivalent,  due to  the following formula for  
indefinite integral (antiderivative): 
\begin{eqnarray}\label{Kin2}
 \int dx \, \pi x \, \cot(\pi x)=
x \log(1-e^{2\pi i x})-\frac{i}{2\pi}
\left( \pi^2 x^2+Li_2(e^{2\pi i x}) \right)\,. 
\end{eqnarray}

Writing functions in the r.h.s. of (\ref{Kin}) or (\ref{Kin2}), (\ref{Kin0}) as a sums over powers of $e^{2\pi i N}$:

\begin{eqnarray}
Li_2(e^{2\pi i N})&=&\sum_{k=1}^{\infty} \frac{e^{2\pi i k N}}{k^2}  \\
N \log\frac{\pi}{\sin \pi N}&=& N\log 2\pi - \frac{i\pi}{2}N+i\pi N^2+N \sum_{k=1}^{\infty}\frac{e^{2\pi i k N}}{k}
\end{eqnarray}
we get:
\begin{eqnarray}
\log \frac{G(1+N)}{G(1-N)}=  \frac{i}{2\pi}\sum_{k=1}^{\infty}\frac {e^{2\pi i kN}}{k^2}+N \sum_{k=1}^{\infty}\frac {e^{2\pi i kN}}{k} +\\ \nonumber
\frac{i\pi}{2}N^2 +N \log 2\pi- \frac{i\pi}{12}
\end{eqnarray}
which coincides with (\ref{GNGN}) due to $\sum_{k=1}^{\infty}\frac {1}{k^2}=\frac{\pi^2}{6} $.

\section{Conclusion}

We conclude, that volume of $SU(-N)$ isn't given by analytically continued volume of $SU(N)$. This is in correspondence with the fact, that volume of $SU(N|M)$ doesn't analytically depend on $N-M$ \cite{Voron}. However, $N \leftrightarrow -N$ remains symmetry of the theory,  realized in more complicated way - volume function give rise to two analytical functions, which combine into the doublet of this symmetry. Similar considerations are applicable to Vogel's symmetry with respect to permutations of parameters. Let's stress that according to  this picture (and this is our general understanding),  $N$ and $-N$ are on the completely equal footing, as well as Vogel's parameters and their any permuted set. Each statement, feature, etc. for a given $N$ (or for given set of Vogel's parameters), has its counterpart for $-N$ (or for permuted set of parameters). 

For full consideration of dependence of analytical volume functions (and Chern-Simons partition functions) on its parameters one need an understanding of analytical properties of Barnes' multiple gamma functions as an analytical functions of parameters. Some initial considerations are given in Sections \ref{reflect}, \ref{reflectsun}, where we calculated the change of volume functions under permutation of Vogel's parameters and make contact with Kinkelin's reflection relation on Barnes' $G$-function. One can continue this line by considerations of branching around zero in the complex plane of each parameter, considerations of analytic properties (with respect to the parameters) of the special combinations of multiple gamma functions, such as multiple sine functions, etc. This last case is relevant for full Chern-Simons partition function on three dimensional sphere. We hope to consider these problems elsewhere. 

The reasonable analogy for the anomaly of Vogel's permutation symmetry seems to be the behavior of  partition functions of some gauge theories under modular transformations of their couplings, \cite{VW94}. As discussed in \cite{W95} in the most simple example of Maxwell theory,  there are two parameters - theta angle $\theta$ and electromagnetic coupling $g$. Theory is unchanged under shift of $\theta \ $ and electromagnetic duality $g^2 \sim 1 / g^2$, which together combine into modular parameter $\tau= \frac{\theta}{2\pi}+\frac{4\pi i}{g^2}$ with usual modular transformation rules. Partition function, however, behaves as modular form, i.e get an additional multiplier, besides the change of arguments. It is interesting to study how far this analogy is going, particularly, whether Vogel's anomaly restricts couplings of the theory with some other fields.

\section{Acknowledgments.}
We are grateful to T.Voronov for encouraging discussions.
We are indebted to MPIM (Bonn), where this work is done, 
for hospitality in autumn - winter 2015-2016. 
Work of RM is partially supported by Volkswagen Foundation 
and by the Science Committee of the Ministry of Science 
and Education of the Republic of Armenia under contract  15T-1C233.

\end{document}